\documentclass[preprint,12pt]{elsarticle}

\usepackage{amssymb}
\usepackage{amsmath}
\usepackage{float}
\usepackage{enumitem}
\usepackage{hyperref}

 \def\btheta{{\boldsymbol {\theta}}}
 \def\bbeta{{\boldsymbol {\beta}}}
  \def\balpha{{\boldsymbol {\alpha}}}
 \def\f{{\tt f}}
 \def\F{{\tt F}}
 \def\k{{\tt k}}
 \def\PP{{\mathbb P}}
 \def\EE{{\mathbb E}}
 \def\Y{{\mathcal Y}}
 
 \usepackage{xcolor}
 
\begin{document}

\begin{frontmatter}


\title{Modelling Stochastic Inflow Patterns to a Reservoir with a Hidden Phase-Type Markov Model}

	\author[ugr]{M.L. Gamiz} 
	
	\affiliation[ugr]{organization={Universidad de Granada},
		country={Spain}}

\author[uja]{D. Montoro} 

\affiliation[uja]{organization={Universidad de Jaen},
	country={Spain}}

\author[ugr]{M.C. Segovia-García} 


\begin{abstract}
	This paper presents a novel methodology for modelling precipitation patterns in a specific geographical region using Hidden Markov Models (HMMs). Departing from conventional HMMs, where the hidden state process is assumed to be Markovian, we introduce non-Markovian behaviour by incorporating phase-type distributions to model state durations. The primary objective is to capture the alternating sequences of dry and wet periods that characterize the local climate, providing deeper insight into its temporal structure.	
	Building on this foundation, we extend the model to represent reservoir inflow patterns, which are then used to explain the observed water storage levels via a Moran model. The dataset includes historical rainfall and inflow records, where the latter is influenced by latent conditions governed by the hidden states. Direct modelling based solely on observed rainfall is insufficient due to the complexity of the system, hence the use of HMMs to infer these unobserved dynamics.
	This approach facilitates more accurate characterization of the underlying climatic processes and enables forecasting of future inflows based on historical data, supporting improved water resource management in the region.

\end{abstract}


\end{frontmatter}

\section{Introduction}
\label{sec:intro}

In recent years, the impact of climate change has added uncertainty to the ever-present uncertainty in water management mechanisms, which aim to be efficient and ensure a reliable supply of this vital resource for human consumption, agriculture, energy production, and industry. In this mission, the application of stochastic models emerges as an indispensable tool. These models allow us to explain the inherent variability in water sources due to the random nature of the hydrometeorological factors involved, such as rainfall events, temperature, evaporation, etc. In addition, they can provide valuable information about the fluctuations in the volume of stored water, which is clearly dependent on the quantities entering (inflows) and leaving (evaporation, water requirements for use, etc.) over time. This knowledge is therefore essential for those in charge of water management, as it plays a crucial role in preventive planning and the efficient use of available resources, as well as in the construction of new structures if necessary. The aim is to avoid the failure of the water supply system and the environmental and socio-economic consequences that this may entail.\\

Stochastic processes are frequently applied in hydrological research to evaluate dam safety under hydrological conditions, the main risk factor being the arrival of intensive rainfall or floods. For example, the authors in \cite{Gabriel-Martin19-2} consider a stochastic model to capture the main hydrometeorological factors involved in dam overtopping reliability; the work in \cite{Gabriel-Martin19} deals with the influence of considering variable reservoir levels prior to the arrival of floods, introducing probability associated to gate failure scenarios. In contrast, the investigation of a potential failure of water management to satisfy use requirements is little addressed by researchers in the field, despite the fact that such a failure is more likely to occur than a failure regarding dam safety, and both can have devastating consequences. Thus, a probabilistic analysis of precipitation is relevant not only to generate knowledge about the arrival of rain events with a view to the safety of the dams, but also to anticipate information to those responsible for the management of the reservoirs in order to ensure an adequate supply of water.\\

Markov processes have shown great advantages from an analytical and practical point of view. They are based on the idea that the state to be occupied in the future does not depend on those occupied in the past but only on the present state. This assumption has traditionally been adopted in the study of precipitation when modeling sequences of wet and dry days or periods (\cite{Feyerman67}, \cite{Gates76}, \cite{MartinVide1999}). Regarding this, a recent work \cite{Channpisey2021} proposes a multi-state model with states representing rainfall intensity levels; the transition probabilities between wet states are governed by a Gamma distribution, that often plays this role together with other distributions such as Pareto, Lognormal or Weibull \cite{Papalexiou2013}. For other purposes of application, the author in \cite{Limnios89} develops a discrete-time Markov model to predict changes in the volume of water stored in a dam reservoir, where reservoir emptying and overflow events are considered as failure states in the dam reliability analysis conducted. Also authors in \cite{Channpisey2021} consider the water storage problem, in this case an optimization problem is addressed to establish the most appropriate water release policy for a rainwater harvesting (RWH) system.\\

Although Markov processes have proven to be widely applicable, they assume exponential sojourn times in their states that may not always fit the observed reality, as discussed in \cite{Thomas}.  Semi-Markovian processes weak such hypothesis and are presented as an immediate extension of Markov processes. They establish that the transition from a state to another depends not only on the present state but also on the time spent in it, which can be governed by any distribution. At this point, Phase-type (PH) distributions emerge as a powerful tool for approaching both types of processes, as they are a dense family in the set of distributions defined in $\mathbb{R}^+$ \cite{Neuts81}. These distributions have been widely applied in the context of reliability and maintenance system modeling [\cite{GamizMoMA2022}, \cite{GamizMoMA2024}, \cite{Li2019}], and as far as the authors are aware, there are not many applications of these distributions in the field of hydrology. Among them, \cite{Ramirez-Cobo} considers a Markovian arrival process (MAP) for rainy day events, and the amount of daily precipitation is PH-distributed. \\

On the other hand, Hidden Markov models (HMM) are of growing interest in practical applications, a recent review can be found in \cite{Mor21}. They model systems in which the state is not directly observable, but is inferred from indirect observations. They are increasingly used in the modeling of natural phenomena; for instance, \cite{Votsi2013} and \cite{Bountzis2020} consider that unobservable ground states can determine the arrival rate and magnitude of earthquakes. Examples of their applications in climatology are \cite{Aillot09}, \cite{Alasseru04}, \cite{Bellone00}, \cite{Euan24}, \cite{Holsclaw16}, \cite{Hughes99} and \cite{Kroiz20}.\\

In this paper we follow similar arguments as in \cite{Alasseru04} where the authors consider a semi-Markov chain of two states: 
‘rain’ and ‘inter-rain’. A semi-Markov model is used to enable every possible duration distribution for ‘rain’ and ‘inter-rain’ events. As a different attempt to address the generalization from exponentially distributed state duration times we consider in this paper the family of phase-type distributions. Here, we propose an HMM model for modeling the annual inflows in a dam reservoir located in Jaen, a region of Andalusia (Spain). Our method lies on the assumption that the inflows distribution is conditioned to the occupation over time of non-observable states (regimes). These states can be interpreted as representations of underlying atmospheric conditions, leading to scenarios that are more or less prone to rainfall and water harvesting. The sojourn times in each of the regimes are assumed to have a discrete phase-type distribution (PH-distribution). Starting from the observation of annual inflows over a period of $N$ years, and using the described approach, the estimation of the model will generate useful information on the pattern of occupancy of the mentioned regimes over time in order to determine the distribution of annual inflows. The forecast of future values of this variable will also provide relevant information for those responsible for water management. The problem involves estimating the parameters of the hidden process as well as the law that produces the observations \cite{Gamiz23}. Due to missing data, a closed-form solution for the MLE is not possible, requiring numerical methods like the EM-algorithm  \cite{Rabiner89}. Theoretical studies include \cite{Baum66} and \cite{Bickel98}.\\

As relevant contributions of this work, we list the following: 1) the use of phase-type distributions allows for the approximation of any general distribution defined in ${\mathbb R}^+$, preserving the Markov property and subsequent advantages in the process. Moreover, well-known distributions such as the Exponential, the Hyperexponential, the Erlang, etc., are particular cases of this family of distributions. 2) The use of HMM models in the field of reservoir management is limited and can provide valuable insights into the behavioral patterns of hydrological variables, taking into account the existence of unobservable factors that interfere with them. This work has focused on modeling the annual inflows to the reservoir; however, in the near future this study could be extended to include the study of evaporation over time or the demand for potable human or irrigation, these being the main sources of water outflow (release) in a reservoir. Thus, anticipating possible changes in the behavioral patterns of the aforementioned variables, perhaps as a consequence of climate change, will allow for appropriate adaptation. \\

The remainder of this paper is organized as follows: First, in Section 2, we describe the Hidden Phase-type Markov Model (PH-HMM) that we consider to model rainfall. Then, in Section 3, we describe the construction of an HMM to represent the behavior of our proposed PH-HMM model. After that, in Section 4, we describe how to estimate the parameters of our model based on the observations of daily rainfall and the EM algorithm. In Section 5, we consider the Moran reservoir model that is dependent on rainfall events and describes the probability that the dam reservoir is at a certain level in a determined year. Considering the Moran model and our PH-HMM model, we obtain some reliability measures in terms of the quantity of water stored in the dam. To showcase our approach, we present a real case study in Section 6, the case of the Quiebrajano reservoir in the province of Jaen (Spain). Finally, we provide some conclusions and future work.\\

	\section{The Hidden Phase-type Markov Model}\label{sec:model}
\subsection{Model Description}
We consider the following parametrization of the model. Let $ \{X_n; n \geq 0\} $ be a discrete-time semi-Markov chain with finite state space $ E = \{1, \ldots, d\} $ and initial law ${\bbeta}=(\beta_1, \ldots, \beta_d)$. We assume that the sojourn times in states follow Phase-type distributions, so we consider a special case of semi-Markov model here. Let $\Y\subseteq {\mathbb R}^+$ a subset of non-negative real numbers, and  $ \{Y_n; n \geq 0\} $ be a $ \Y $-valued sequence such that given $ X_n = i $, $ Y_n $ is conditionally distributed with density $ g(i,y) $ with respect to some $ \sigma $-finite measure $ \mu $ on $ \Y $, and fixed $ i \in E $.

All processes are defined on a complete probability space $ (\Omega, \mathcal{F}, P) $, and this will imply also a family of probabilities $ \{P_i; i \in E\} $, where $ P_i(\cdot) = \PP(\cdot | X_0 = i) $, and expectations $ \{E_{(i,y)}; i \in E, y \in Y\} $, where $ E_{(i,y)}[\cdot] = \EE_{(i,y)}[\cdot | X_0 = i, Y_0 = y] $.

For a specific state of the chain $ \{X_n, n\ge 0\} $, $ i \in E $, let us denote $ \tau_i $ the corresponding sojourn time. We assume for $ \tau_i $ a discrete phase-type distribution. More specifically, $ \tau_i $ behaves as the absorption time of a discrete-time Markov chain with $ F_i $ transient states and one absorbing state. We denote the initial law as $ \balpha_{i} = (\alpha_{(i,1)}, \ldots, \alpha_{(i,\F_i)}, 0) $, and the transition matrix is given by:

\[
{\bf T}_i = 
\begin{pmatrix}
	T_i & T_i^0 \\
	0 & 1
\end{pmatrix},
\]

where $ T_i $ governs the changes between transient states and $ T_i^0 $ is the absorption vector. The usual notation is $ DPH(\alpha_i, T_i) $, for PH-distributions.

All the initial vectors $ \balpha_1, \ldots, \balpha_d $, all matrices $ {\bf T}_1, \ldots, {\bf T}_d $, as well as the family of densities $ \{g(i, \cdot); i \in E\} $, depend on a vector of parameters $ \theta $, that is, $ T_i = T_i(\theta) $ and $ g(i, \cdot) = g_\theta(i, \cdot) $, for all $ i \in E $. The set of possible values of the vector $ \theta $ is denoted $ \Theta \subset \mathbb{R}^+ $, and $ \theta $ has to be estimated from a set of observations of the process $ \{Y_n, n\ge 0\} $.

The vector $ \theta $ usually includes the parameters of the PH-distributions defined above and also some parameters characterizing the densities $ g $.

In general, if $ Y_n $ has density function $ g(i, \cdot) $, given that $ \{X_n = i \}$, then, for  any  $B \subset \mathbb{R}$,
\[
\PP(Y_n \in B | X_n = i) = \int_B g(i,y) \mu(dy), 
\]
is the probability that $ Y_n $ takes values in a subset $ B $ given the event $ \{X_n = i \} $, for $ i \in E $. We will denote $ G(i, B) = \PP(Y_n \in B | X_n = i) $.

For the simpler case that $ Y_n $ takes values in a finite set, that is, $ \Y = \{y_1, \ldots, y_s\} $, then the emission function will be a $ (d \times s)  $-dimensional matrix $ \bf G $, with elements 
\[
g(i,y) = G(i, \{y\}) = \PP(Y_n = y | X_n = i), 
\]
for all  $i \in E $ and  $y \in \Y$,  and all  $n \ge  0$. 

For the rest of the paper, we will consider the process $ (X,Y) $ defined as follows.

\subsection*{Definition 1: The PH-HMM Model}
Let $ (X,Y) = \{(X_n,Y_n); n \geq 0\} $ be a two-dimensional process with dependence structure $ M_1 - M_0 $, where $ X_n $ is an irreducible homogeneous semi-Markov process in a finite set $ E $, with initial distribution $ \bbeta $ and with sojourn time in state $ i $ given by a DPH-distribution with parameters $ (\balpha_i, T_i) $. Then the kernel matrix of $ X $ is a $ d- $dimensional matrix with entries 

\[
Q_{ij}(n) = p_{ij} F_{ij}(n),
\]
where 
\[
F_{ij}(n) = \balpha_i (T_i^{(n-1)} T_i^0 \otimes \balpha_j)e,
\]
and $ p_{ij} = Q_{ij}(\infty) $ is the probability of a jump from state $ i $ to $ j $. Moreover, $ Y_n $ is a homogeneous process in a finite set $ \Y = \{y_1, \ldots, y_s\} $, such that, for $ n \geq 0 $, the distribution of $ Y_n $ is determined by $ g(i, \cdot) $ over the event $\{ X_n = i \}$, for $ i \in E $.

We refer to the states of the hidden process as regimes, while the values of the observable process are called signals. The matrix $ {\bf G} = (g(i,y); i \in E, y \in \Y) $, defined above is called the emission matrix. 

\begin{figure}[t]
	\centering
	\includegraphics[width= 0.95\textwidth]{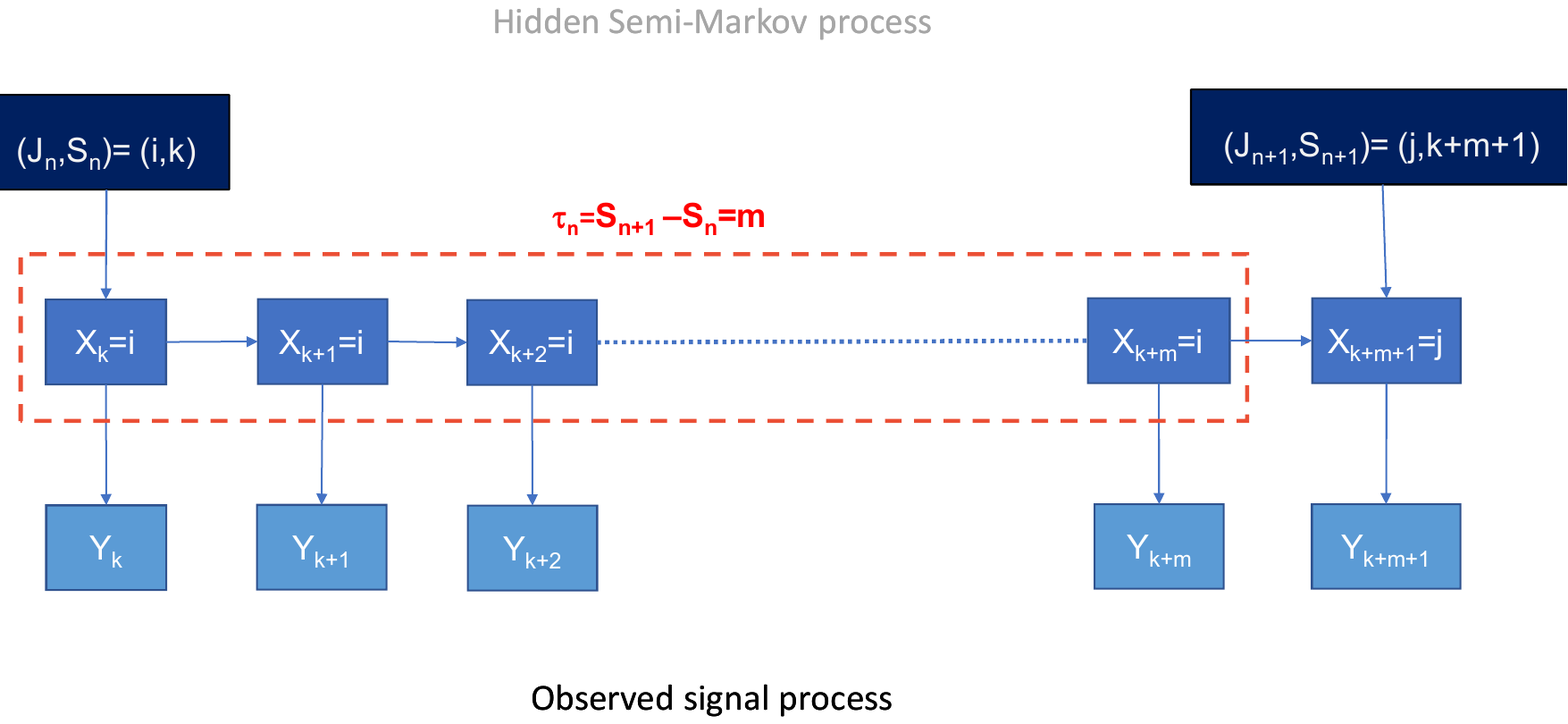}
	\vspace{-2cm}
	\caption{A discrete time hidden Markov process evolution}\label{fig:model}
	\end{figure}

 Let $\{J_n, n\ge n\}$ denote the sequence of visited states, and $\{S_n, n\ge n\}$ the successive sojourn times, then $(J,S)=\{(J_n,S_n), n\ge 0\}$ is a  Markov renewal process (MRP), that is, we assume that
\begin{eqnarray*}
&&\PP(J_{n+1}=j, S_{n+1}-S_n=m | (J_1,S_1),\ldots (J_n=i,S_n))=\\
&&\qquad \PP(J_{n+1}=j, S_{n+1}-S_n=m | J_n=i)
\end{eqnarray*}
for all $i, j \in E$, and all $n\ge 0$. Let $N_k=\max\{n:S_n\leq k\}$, then the process $X=\{X_k, k\ge 0\}$, where $X_k=J_{N_k}$ is a semi-Markov process associated to the MRP $(J,S)$. A graphical representation  of the DTHSMM model evolution with time is represented in Figure \ref{fig:model}. At the $k$-th transition time, the process enters state $i$, which the $n$th visited state (i.e. $n\leq k$). The process remains in state $i$ for a total of $m$ steps, and at the time $k+m$ it jumps to a different state $j$. Each time $k$ a signal $Y_k$ is emitted.   

\subsection*{Practical case: A PH-HMM to Model Yearly Rainfalls}
\noindent Let us consider $ Y_n $ representing the total amount of rain registered in a particular area during the $ n $-th year, with $ n = 0, 1, \ldots $. It is known that the amount of rainfall in a particular area can depend on a variety of factors, including: Atmospheric Circulation Pattern (ACP), topography, proximity to bodies of water, temperature, humidity levels, elevation, land use, and, of course, climate change. So there is certainly some kind of non-observable heterogeneity present in the distribution of $ Y_n $. 

We assume that all these factors can be summarized by a latent random variable $ X_n $ in such a way that, given  $ \{X_n = i\} $, the probability of an event $ \{Y_n \in B\} $ can be determined independently of the past.

In our simplest model, we assume the state space of the chain $ X_n $ to be $ E = \{1,2\} $. That is, two different regimes are considered under which the rainfalls evolve. Without any loss of generality, we can consider, for example, that state 1 represents the regime more prone to rain, while state 2 is more associated with dry seasons. 

It is reasonable also to assume that the duration of a wet regime, $ \tau_1 $, and the duration of a dry regime, $ \tau_2 $, are of very different nature. Also, the non-memory property might not be realistic for $ \tau_1 $ and/or $ \tau_2 $, that is, hazards are not presumably constant with time, and then the Exponential law might not be appropriate. To be more realistic, we consider for each regime duration a PH-distribution \cite{Neuts81}.

\section{Solving PH-HMMs by means of HMM}\label{sec:solving}
In this section, we will construct an HMM $ \{( \widetilde{X}_n, Y_n )\} $ that properly represents the behaviour of the PH-HMM $ \{(X_n, Y_n)\} $ with parameter vector $ \theta $ defined in the previous section. To do it, the extended state-space strategy will be used as explained next.

For $ i = 1, \ldots, d $, let us assume that the duration of regime $ i $ has a discrete PH-distribution with parameters $ (\alpha_i, T_i) $. Let us denote $ F_i $ the order of this DPH, that is, the corresponding number of phases, which is the dimension of matrix $ T_i $. For simplicity, we assume that $ \F_i = \F $ for all $ i = 1, \ldots, d $.

We define the following set of ‘regime-phase’ states,
$$ \widetilde{E} = \{(1,1), \ldots, (1,\F), \ldots, (d,1), \ldots, (d,\F)\} $$
 with cardinality $ \widetilde{d} = d \cdot \F $. Define $ \widetilde{X}_n $ a random variable that takes the value $ (i,\f) $ when $ X_n = i $ and the sojourn time in $ i $ is running under the phase $ \f $. Then $ \{\widetilde{X}_n\} $ is the Markov chain with state space $ \widetilde{E} $ and transition matrix

\[
\widetilde{\bf P} =
\begin{pmatrix}
	T_1 & p_{12} T_1^0 \otimes \balpha_2 & \cdots & p_{1d} T_1^0 \otimes \balpha_d \\
	p_{21} T_2^0 \otimes \balpha_1 & T_2 & \cdots & p_{2d} T_2^0 \otimes \balpha_d \\
	\vdots & \vdots & \ddots & \vdots \\
	p_{d1} T_d^0 \otimes \b alpha_1 & p_{d2} T_d^0 \otimes \balpha_2 & \cdots & T_d
\end{pmatrix},
\]

and initial distribution 

\[
\widetilde{\bbeta} = (\beta_1 \cdot \balpha_1, \ldots, \beta_d \cdot \balpha_d),
\]

a vector of dimension $ \widetilde{d} $. The matrix $\widetilde{\bf P}$ is composed of submatrices of dimension $d \times d$, each.

\subsection*{Remark 1: Assumption}
The signal process $ \{Y_n, n\ge 0\} $ depends on the state of $ X_n $, but not on the particular phase the corresponding sojourn time is in, then 

\[
\PP(Y_n = y | \widetilde{X}_n = (i,\f)) = \PP(Y_n = y | \widetilde{X}_n = (i,k)) = \PP(Y_n = y | X_n = i).
\]

Using this transformation of semi-Markov chain $ \{X_n\} $ into the Markov chain $ \{\widetilde{X}_n\} $, we can transform the PH-HMM $ \{(X_n,Y_n)\} $ into the HMM $ \{(\widetilde{X}_n,Y_n)\} $, with parameter vector $ \widetilde{\theta} $ determined by the transition matrix of the MC $ \{\widetilde{X}_n\} $, that is $ \widetilde{\bf P} $, with entries

\[
\widetilde{P}((i,\f),(j,\k)) = T_i(\f,\k) \delta_{ij} + T_i^0(\f) \alpha_{(j,\k)}(1 - \delta_{ij}),
\]
for all $ i,j \in E $, $ \f,\k = 1, \ldots, \F $, and emission matrix $ \widetilde{\bf G} $, with entries

\[
\widetilde{G}((i,\f),y) = g(i,y)
\]
\section{Estimation of the extended HMM}\label{sec:estima}
Let us assume that we have a sample of observations $ Y_1^N = \{Y_1, \ldots, Y_N\} $, from which we want to infer appropriate values for the unknown vector of parameters $ \theta $. With complete data, that is, with observations $ \{(\widetilde{X}_1, Y_1), \ldots, (\widetilde{X}_1, Y_N)\} $, we would build the full log-likelihood function as

\begin{eqnarray*}
l(\theta) &=& \sum_{n=1}^N \sum_{(i,\f)} \sum_{(j,\k)} 1_{\{\widetilde{X}_n = (i,\f), \widetilde{X}_{n+1} = (j,\k)\}} \log \widetilde{P}((i,\f),(j,\k)) + \\
&&\qquad \sum_{n=1}^N \sum_{i \in E} \sum_{\f=1}^\F 1_{\{\widetilde{X}_n = (i,\f)\}} \log g(i,Y_n).
\end{eqnarray*}

We cannot directly optimize this expression since we do not have complete data. Instead, we use the EM algorithm. At the $ r $-th iteration of the algorithm, we have an estimation $ \hat{\theta}^{(r)} $, and define

\[
{\theta}^{(r+1)} = \underset{\theta}{\arg\max} \left\{ \psi(\theta | {\theta}^{(r)}) \right\},
\]

where 

\[
\psi(\theta | {\theta}^{(r)}) = \psi_1(\theta | {\theta}^{(r)}) + \psi_2(\theta | {\theta}^{(r)}),
\]

with

\[
\Psi_1(\theta | {\theta}^{(r)}) = \sum_{n=1}^N \sum_{(i,j)} \sum_{(\f,\k)} \PP_{{\theta}^{(r)}}(\widetilde{X}_{n-1} = (i,\f), \widetilde{X}_n = (j,\k) | Y_1^N) \log \widetilde{P}((i,\f),(j,\k)),
\]

and

\[
\Psi_2(\theta | {\theta}^{(r)}) = \sum_{n=0}^N \sum_{i \in E} \sum_{\f=1}^\F \sum_{y \in Y} \PP_{{\theta}^{(r)}}(\widetilde{X}_n = (i,\f)) 1_{\{Y_n = y\}} \log g(i,y).
\]

\subsection{M-step}
On the one hand, direct maximization of expression (1) leads to

\[
\widehat{\widetilde{P}}((i,\f),(j,\k)) = \frac{\sum_{n=1}^N \PP_{{\theta}^{(r)}}(\widetilde{X}_{n-1} = (i,\f), \widetilde{X}_n = (j,\k))}{\sum_{n=1}^N \PP_{{\theta}^{(r)}}(\widetilde{X}_{n-1} = (i,\f))},
\]

for $ i,j \in E $, and $ \f,\k = 1, \ldots, \F $. On the other hand, maximizing expression (2), we get

\[
\widehat{G}(i,y) = \frac{\sum_{n=1}^N P_{{\theta}^{(r)}}(X_{n-1} = i) 1_{Y_n = y}}{\sum_{n=1}^N P_{{\theta}^{(r)}}(X_{n-1} = i)}.
\]

\subsection{E-step}
Let us define the following family of probabilities,

\[
F^{(r)}_n(i,\f) = \PP_{{\theta}^{(r)}}(Y_1^n, \widetilde{X}_n = (i,\f)),
\]

and

\[
B^{(r)}_n(i,\f) = \PP_{{\theta}^{(r)}}(Y_{n+1}^N | \widetilde{X}_n = (i,\f)),
\]

for all $ n = 1, \ldots, N; i,j \in E; \f = 1, \ldots, \F $. Additionally, we define $ F_0(i,\f) = \beta_i \alpha_{(i,\f)} $, and $ B_N(i,\f) = 1 $, for all $ i \in E, \f = 1, \ldots, \F $. Using the forward-backward algorithm, we have that

\[
F_n(i,\f) = \sum_{j \in E} \sum_{\k=1}^\F F_{n-1}(j,\k) \widetilde{P}((j,\k),(i,\f)) g(i,Y_n),
\]
and
\[
B_n(i,\f) = \sum_{j \in E} \sum_{\k=1}^\F \widetilde{P}((i,\f),(j,\k)) g(j,Y_{(n+1)}) B_{n+1}(j,\k),
\]
for all $ n = 1, \ldots, N-1 $, where we have omitted superscript ${(r)}$ for simplicity in notation.

Using these families of probabilities, we can solve the E-step at the $ r $-th iteration of the algorithm. That is,

\[
\PP_{{\theta}^{(r)}}(\widetilde{X}_n = (i,\f) | Y_1^N) = \frac{F_n^{(r)}(i,\f) B_n^{(r)}(i,\f)}{\PP_{{\theta}^{(r)}}(Y_1^N)},
\]
and
\[
\PP_{{\theta}^{(r)}}(\widetilde{X}_{n-1} = (i,\f), \widetilde{X}_n = (j,\k) | Y_1^N) = \frac{F_{n-1}^{(r)}(i,\f) \widetilde{P}^{(r)}((i,\f),(j,\k)) G^{(r)}(j,Y_n) B_n^{(r)}(j,\k)}{\PP_{{\theta}^{(r)}}(Y_1^N)}.
\]

Once convergence is reached, we can obtain estimations of the elements of vector $ \theta $ from $ \widetilde{P}^{\hat{}} $ as follows:

For all $ i \in E $, we get the transition probabilities between phases by means of:

\[
\widehat{T}_i(\f,\k) = \widehat{\widetilde{P}}((i,\f),(i,\k)),
\]

and then 

\[
\widehat{T}_i^0 = ({\bf I} - \widehat{T}_i){\bf e}.
\]

The initial law of the PH-distribution for state $ j $ can be deduced from the equation 

\[
\widetilde{P}((i,\f),(j,\k)) = T_{(i,\f)}^0 p_{ij} \alpha_{(j,\k)},
\]

using the restrictions 

\[
\sum_{j \neq i} p_{ij} = 1, \quad \text{and} \quad \sum_{\k=1}^\F \alpha_{(j,\k)} = 1, \quad \text{for all } i, j \in E.
\]

The estimations of the emission probabilities $ g(i,y) $ are obtained directly for all $ i \in E $, and $ y \in Y $.

\section{The Moran Reservoir Model}
The Moran reservoir model strongly depends on the rain events probability distribution, which we assume as the only supply source for the reservoir \cite{Limnios89}. In the previous sections, we have presented a new rainfall model based on a hidden semi-Markov chain structure. That is, we have used a coupled stochastic process. The first component models the “rain” and “inter-rain” states using a special case of semi-Markov process, where the sojourn time in each state is modeled by a DPH-distribution. The second component accounts for the amount of water collected in a pre-specified period of time (daily, monthly, or yearly, depending on the particular interest).

\subsection{Description of the Model}
Following the procedure explained in the previous sections, we can derive the marginal distribution of the rain series as follows. For $ n \ge 0 $,

\[
\PP(Y_n \in B) = \sum_{i \in E} \PP(Y_n \in B | X_n = i) \beta_i
\]

for any subset $ B \subset Y $. 

Consider a discrete reservoir with capacity $ C $ in volume units. We assume that the only supply is due to rain represented by a family of random variables $ \{Y_n; n \geq 0\} $, with distribution determined as explained above.

Let $ V_n $ denote the reservoir level at the $ n $-th time (year). During the time period $ (n, n+1) $, a quantity of water $ Y_n $ enters the reservoir, and during this interval, a fixed output flow equal to $ \omega $ units leaves the reservoir, if available. If at the $ n $-th time the input flow $ Y_n $ is greater than $ C - V_n $, a quantity $ Y_n - (C - V_n) $ is lost as overflow. Hence,

\[
V_{n+1} = 
\begin{cases}
	0, & \text{if } V_n + Y_n < \omega \\
	V_n + Y_n - \omega, & \text{if } \omega \leq V_n + Y_n < C \\
	C - \omega, & \text{if } V_n + Y_n \geq C
\end{cases}
\]

Then $ \{V_n; n \geq 0\} $ is a Markov chain (see \cite{Limnios89}). Let us represent the state space by the set $ \Omega = \{0, 1\omega, 2\omega, \ldots, n_0\omega, C\} $, with $ n_0 = \left\lfloor \frac{C}{\omega} \right\rfloor $, where $ \left\lfloor \cdot \right\rfloor $ is the integer part of a number. State $ i\omega $ indicates that the volume of water in the reservoir is in the interval $[i\omega, (i+1)\omega)$. If observations are taken yearly, this means that the water supply of the population is guaranteed for at least the next $ i $ years, even if no rain is received. Notice that in general, $ Y_n $ can take values in $ \mathbb{R}^+ \cup \{0\} $.

The transition probability matrix for the MC $ \{V_n; n \geq 0\} $ is given by

\begin{eqnarray}\label{eq:moran}
&&{\bf P }= \\
&&\nonumber \begin{pmatrix}
	\PP(0 \leq Y \leq \omega) & \PP(\omega < Y \leq 2\omega) & \PP(2\omega < Y \leq 3\omega) & \cdots & \PP(Y > n_0\omega) \\
	\PP(Y = 0) & \PP(0 < Y \leq \omega) & \PP(\omega < Y \leq 2\omega) & \cdots & \PP(Y > (n_0-1)\omega) \\
	0 & \PP(Y = 0) & \PP(0 < Y \leq \omega) & \cdots & \PP(Y > (n_0-2)\omega) \\
	\vdots & \vdots & \vdots & \ddots & \vdots \\
	0 & 0 & 0 & \PP(Y = 0) & \PP(Y > 0)
\end{pmatrix}.
\end{eqnarray}

We need to determine the probability $ \PP(Y = y) $, which in principle depends on the time $ n $, that is,

\begin{eqnarray*}
\PP(Y_n = y) &=& \sum_{i \in E} \PP(Y_n = y | X_n = i) \PP(X_n = i) = \sum_{i \in E} g(i,y) \PP(X_n = i)\\
&=&\sum_{i \in E} g(i,y) \sum_{\f =1}^{\F} \PP(\widetilde{X}_n = (i,\f))
\end{eqnarray*}

In matrix notation,

\[
{\bf p}_n^Y = \widetilde{\boldsymbol \pi}_n ({\bf I}_d \otimes {\bf e}) {\bf G},
\]
where $\widetilde{\bf \pi}_n$ is the stationary distribution of the MC $\{\widetilde{X}_n,n \ge 0\}$, ${\bf e}$ is a vector of 1, and $\otimes$ represents Kronecker product of matrices.

One conclusion is that when the MC $ \{\widetilde{X}_n; n \geq 0\} $ is stationary, then the transition probabilities of the MC $ \{V_n; n \geq 0\} $ do not depend on $ n $, so that the matrix $ \bf P $ given above is stationary. When the observation times coincide with the jump times of the SMC $ X_n $, then it is enough for the embedded MC to be stationary to get the transition probabilities of $ V $ to be homogeneous.

\subsection{Evaluation Measures}\label{sec:measures}
We call ‘empty’ the state in which the volume of water in the reservoir is not enough to provide one single year of water demand. In particular, we are interested in the following dependability measures:

\begin{itemize}
	\item {\it Reliability}: \\
	Given an initial state $\{V_0=v\}$ we define $R_v(n)$ as the probability that the reservoir is not ``empty'' at any time before $n+1$, for all $n >0$, that is, 
	\[
	R_v(n) ={\bf 1}_v {\bf P}_0^n{\bf e},
	\]
	with $ {\bf 1}_v $ being a vector of dimension $ n_0+1 $, with zeros except for position $ v $ which is 1; and $ P_0 $ denoting the transition matrix corresponding to the MC $ V_n $, without considering the “empty” state.
	
	\item {\it Availability}: \\
		Given an initial state $\{V_0=v\}$ we define $A_v(n)$ as the probability that the reservoir is not ``empty'' at time before $n$, for all $n >0$, which can be calculated as 
	\[
	A_v(n) = 1_v {\bf P}^n {\bf e},
	\]
	for $n=1,2,\ldots$. In this case, $ {\bf 1}_v $ is a vector of dimension $ n_0+2 $, with zeros except for position $ v $ which is 1; while ${\bf e}$ is a vector of 1 with the appropriate dimension.
	
	\item {\it Mean Time to Failure {\rm (MTTF)}}: \\
	The mean time until the first ``empty'' state is reached can be calculated as
		\[
	\text{MTTF}_v = \sum_{n > 0} n(R_v(n-1) - R_v(n)),
	\]
	for any  $v$ initial state.

	Notice that we can calculate these measures independently of the initial state.
\end{itemize}

\subsection{A Simulation Example}
Consider the following particular conditions for simulation:

\begin{itemize}
	\item Let the SMC $\{X_n\}$ with $E=\{1,2\}$ and $\bbeta=(0.6,0.4)$.
	\item Let the sojourn times in states the ${\tt DPH}$:
	
	{\small{	\[
			\tau_1 \hookrightarrow {\tt DPH} \left(\balpha_1=(0.5,0.5),{\bf T}_1=\left(\begin{array}{cc}0.5&0.4 \\ 0.3 &0.5 \end{array}\right)\right),
			\]
			\[
			\tau_2 \hookrightarrow {\tt DPH} \left(\balpha_2=(0.5,0.5),{\bf T}_2=\left(\begin{array}{cc}0.3&0.3 \\ 0.2 &0.5 \end{array}\right)\right)
			\]}}
	\item The emission distribution:
	\begin{itemize}
		\item $g(1,y)=1$ if $y=0$ and 0 elsewhere, under the event $\{X_n=1\}$; and,
		\item $g(2,y)=dpois(y,\lambda=5)$; when  $\{X_n=2\}$;
	\end{itemize}
	\item For the MC $\{V_n\}$,  take $\omega=5$ and $C=20$.
\end{itemize}

In this case, we are considering a simplified state space for the MC modelling the volume of the reservoir with only 5 states, $ E_V = \{0, 1, 2, 3, 4\} $, with 0 denoting the “empty” or failure state, and $ i $ denotes the state in which the volume of water in the dam is enough for human supplying during at least the next $ i $ years, $ i = 1, 2, 3, 4 $. The true matrix of transition probabilities is given by 

\begin{equation}\label{eq:trueP}
{\bf P} = \begin{pmatrix}
	.881 & .115 & .004 & .000 & .000 \\ 
	.693 & .189 & .115 & .004 & .000 \\ 
	.000 & .693 & .189 & .115 & .004 \\ 
	.000 & .000 & .693 & .189 & .119 \\ 
	.000 & .000 & .000 & .693 & .307 
\end{pmatrix}.
\end{equation}

Under these conditions, we have simulated a total of $ M = 1000 $ samples of size $ N = 100 $. In Figure \ref{fig:measures} (left panel), we present the average of the reliability values estimated along the $ M $ samples at times $ n = 1, \ldots, 10 $. 
Let us consider the following notation: $x_{n}^{n+m}=(x_{n},x_{n+1},\ldots,x_{n+m})$; and $y_{n}^{n+m}=(y_{n},y_{n+1},\ldots,y_{n+m})$, for $n,m \geq 0$.\\
The simulation algorithm can be summarized as follows:

\begin{enumerate}[label=Step \arabic*.] 
	\item Let $N$ be the sample size. Put $n=0$, and choose an initial hidden state $x_0$ from $E=\{1,\ldots, d\}$ according to $\bbeta$. Generate $y_0$ from the distribution law $g({x_0};\cdot)$;
	\item Generate  $t_n$ the duration time in state $x_n$ from the ${\tt DPH}$  of $\tau_{x_n} $. Define $N_0=\min\{N,n+t_n\}$. Put $x_{n+1}^{N_0}=x_n$ ; 
	\item  For $s_n=n+1,\ldots,N_0$, generate $N_0-n+1$ signals from the law $g(x_n;\cdot)$, that is, obtain $y_{n+1}^{N_0}$
	\item  If  $N_0=N$, then stop. Else put $n=N_0+1$, select $x_n$ from $P_{x_{n-1},x_n}$, the vector of jump probabilities and go to Step 2.
\end{enumerate}

\subsection*{Some simulation results}
\noindent We simulate a total of M=1000 samples of size N=100, $\{y_1, y_2,…,y_N\}$, using the above simulation algorithm. For each sample we estimate the parameters of the underlying PH-HMM,using the procedure discussed in Sections \ref{sec:model} and \ref{sec:solving}. Then we follow the arguments described in Section \ref{sec:estima} to estimate, based on each sample, the reliability and the availability of the reservoir as defined previously in Section \ref{sec:measures}. \\

The obtained results are summarized in the following. 
\begin{itemize}
	\item Estimated parameters of the PH-HMM, i.e., $\widehat{\btheta}$:\\
	Based on $M=1000$ samples of size $N=100$ we obtain that
	\[\tau_1 \hookrightarrow {\tt DPH} \left(\widehat{\balpha}_1=(0.603,0.397),\widehat{{ T}}_1=\left(\begin{array}{cc}0.599&0.361 \\ 0.243 &0.369 \end{array}\right)\right),
	\]
	\[
	\tau_2 \hookrightarrow {\tt DPH} \left(\widehat{\balpha}_2=(0.611,0.389),\widehat{ T}_2=\left(\begin{array}{cc}0.202&0.312 \\ 0.230 &0.488 \end{array}\right)\right)
	\]
	which are very close to the true values given above.
	\item The average of the estimated transition probability matrices along the 1000 samples is obtained as 
	\[
	Ave(\widehat{\bf P})=
	\begin{pmatrix}
		
		.880 & .109 & .010 & .001 & .000 \\ 
		.698 & .182 & .109 & .010 & .001 \\ 
		.000 & .698 & .182 & .109 & .011 \\ 
		.000 & .000 & .698 & .182 & .120 \\ 
		.000 & .000 & .000 & .698 & .302 \\ 
	\end{pmatrix}
	\]	
	providing very close values to the probabilities of the true matrix presented in \eqref{eq:trueP}.

	\item \textbf{Reliability measures:}
		\begin{figure}[ht]
		\centering
		\includegraphics[width=0.45\textwidth]{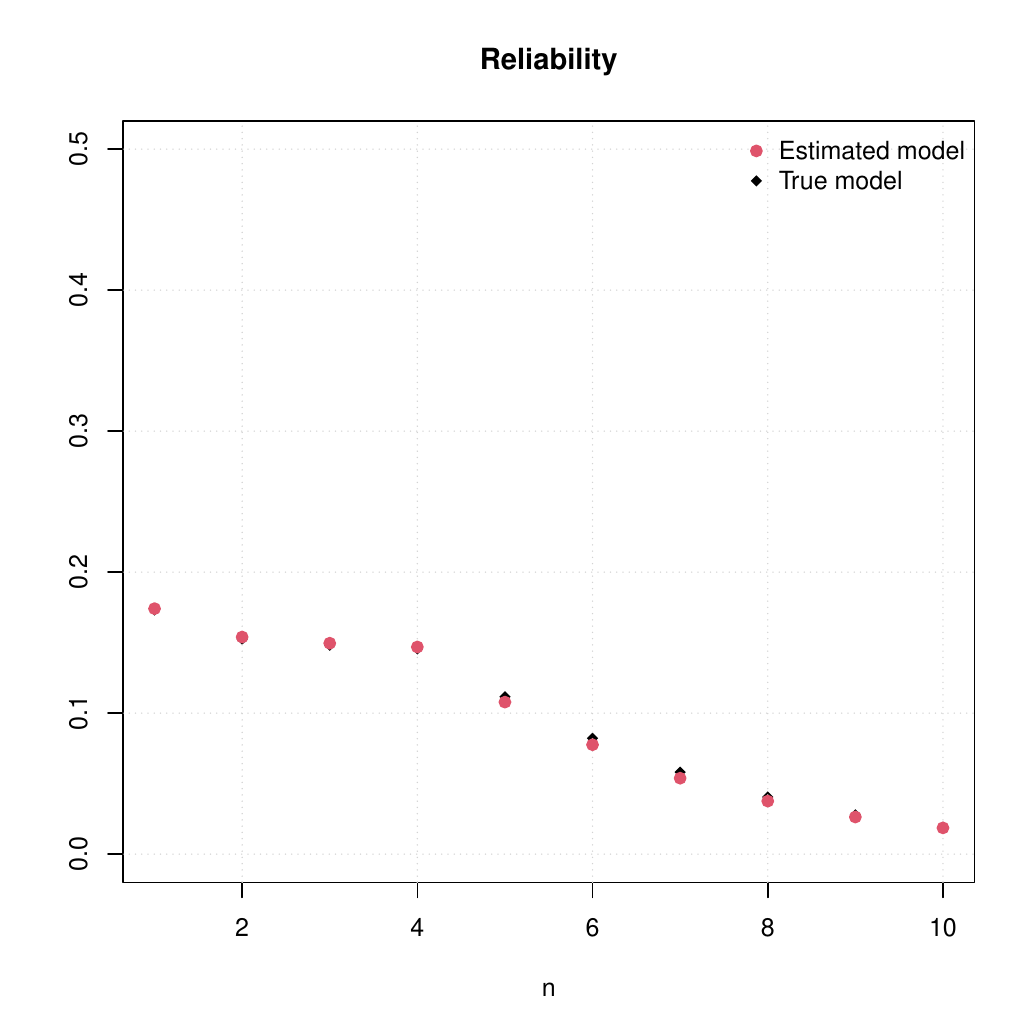}	
		\includegraphics[width=0.45\textwidth]{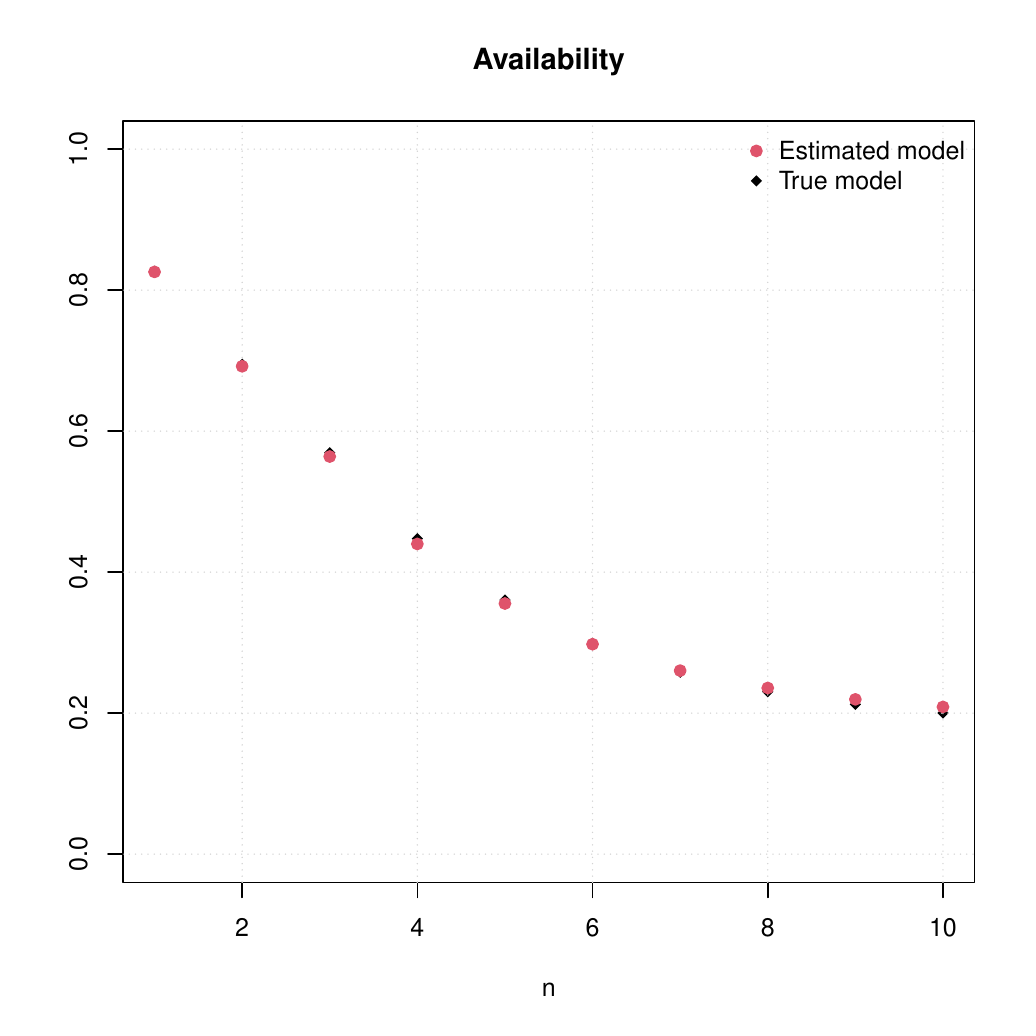}
		\caption{Performance measures}\label{fig:measures}
	\end{figure}

	The estimated values of the reliability and availability measured defined above are presented in Figure \ref{fig:measures}. The red points represent the average values obtained along all the simulated samples while the black points represent the true model. If we interpret our model in the context of a reservoir these estimated performance measures would be interpreted as follows: At time 2, the reliability of the reservoir is low (0.15), meaning there is a high probability that the reservoir has failed to meet water demand at least once by this point. This reflects frequent early-stage operational shortages, likely due to insufficient initial storage, high demand, or low inflows.	
	However, the relatively high availability at time 2 (0.8) indicates that despite these failures, the system is operational in most time steps, thanks to inflow events that restore the reservoir's volume above the critical threshold. In other words, the system frequently recovers between failures, maintaining an acceptable level of service most of the time.
	
	\item 	{\bf MTTF:}\\
	\begin{figure}[ht]
		\centering
		\includegraphics[width=0.6\textwidth]{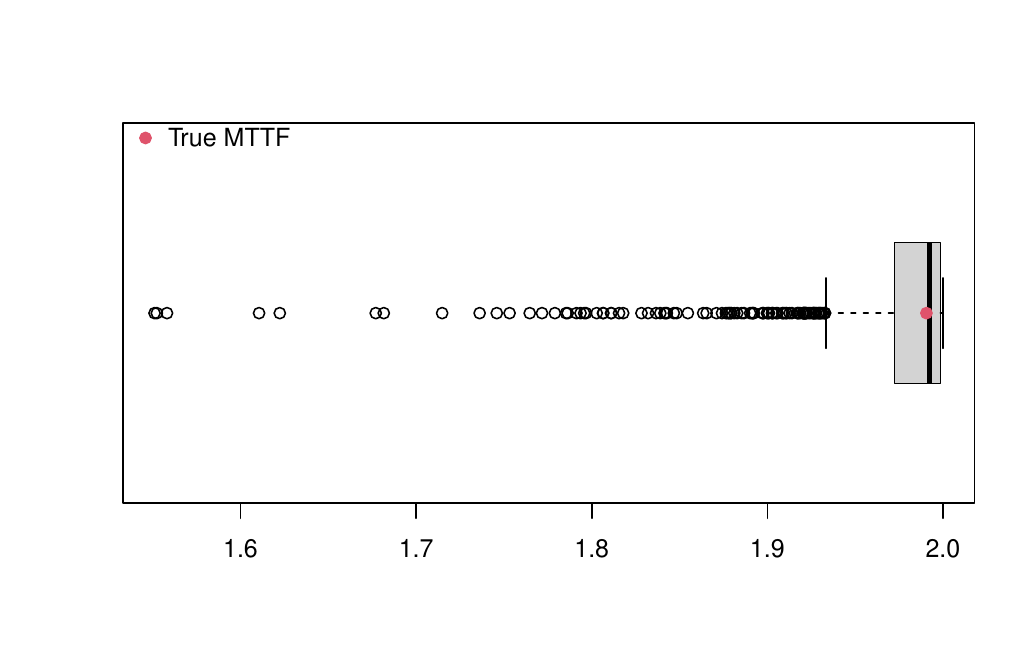}
		\caption{MTTF}\label{fig:mttf}
	\end{figure}
	The average MTTF across the simulated samples is 1.989, while the true MTTF, calculated using the actual parameters, is 1.999. Figure \ref{fig:mttf} presents a boxplot of the estimated mean time to failure from repeated experiments. The MTTF based on the true model is highlighted in red. As can be seen from the graph, the median of the estimated MTTFs provides an accurate approximation of the true MTTF.
	The distribution of MTTF values obtained from the 1000 simulations exhibits a pronounced left skewness, with a small but significant proportion of samples experiencing failure at early times. This asymmetry reflects the presence of extreme early-failure scenarios, likely driven by adverse initial conditions or sequences of low inflow. While the majority of simulations indicate long operating periods before the first failure, in accordance with the true model, these outliers substantially reduce the mean MTTF, resulting in a skewed distribution. Such behaviour is consistent with first-passage processes under stochastic hydrological forcing.
	
	\end{itemize}

\section{A real case study: The Quiebrajano dataset}
The Quiebrajano reservoir, located in the province of Jaen (Spain) and operational since 1976, plays a crucial role in regional water supply.

Based on historical data from the SAIH system (\textit{Sistema Automático de Información Hidrológica}) \cite{saih}, the reservoir releases an average of approximately 0.908 $hm^3$ per month (10.518  $hm^3/year$) through its controlled outflows. The outflow values exhibit high interannual variability, with interquartile range {\bf IQR}= $Q_3-Q_1=47.52- 0.92 = 46.6$ $hm^3$.

These variations reflect changing hydrological conditions, operational decisions, and water demand patterns. With an estimated average inflow of around 12.33  $hm^3/year$ the dam typically operates under a constrained water balance, where outflows represent a significant share of available resources and leave limited margin during dry periods.

The main objective of this section is to illustrate the model presented in Sections 2 and 3 using a real case study. While the SAIH system provides valuable real-time and historical data on key hydrological variables such as reservoir levels, inflows, and outflows, its utility for in-depth scientific analysis is limited by several factors. These include the temporal resolution of the data, the lack of contextual metadata, information on management decisions or upstream conditions, and potential gaps or inconsistencies in long-term records (see Figure \ref{fig:yearvol}).
 Therefore, although the SAIH system is a critical operational tool for basin management and situational awareness, researchers often need to complement its data with other sources to ensure analytical robustness. Consequently, this paper develops a preliminary approach to analyse the reservoir’s hydrological balance by using a phase-type hidden Markov model (PH-HMM) to characterize inflow behaviour, followed by a Moran model to describe the temporal dynamics of the stored volume in the dam.\\
The numerical analyses were performed using custom code along with functions implemented in the following R packages: $HiddenMarkov$ \cite{hiddenmarkov}, $markovchain$ \cite{markovchain}, and $forecast$ \cite{forecast}.

\subsection{Fitting a PH-HMM to explain inflow distribution}
\begin{figure}[H]
\centering
\includegraphics[width=0.9\textwidth]{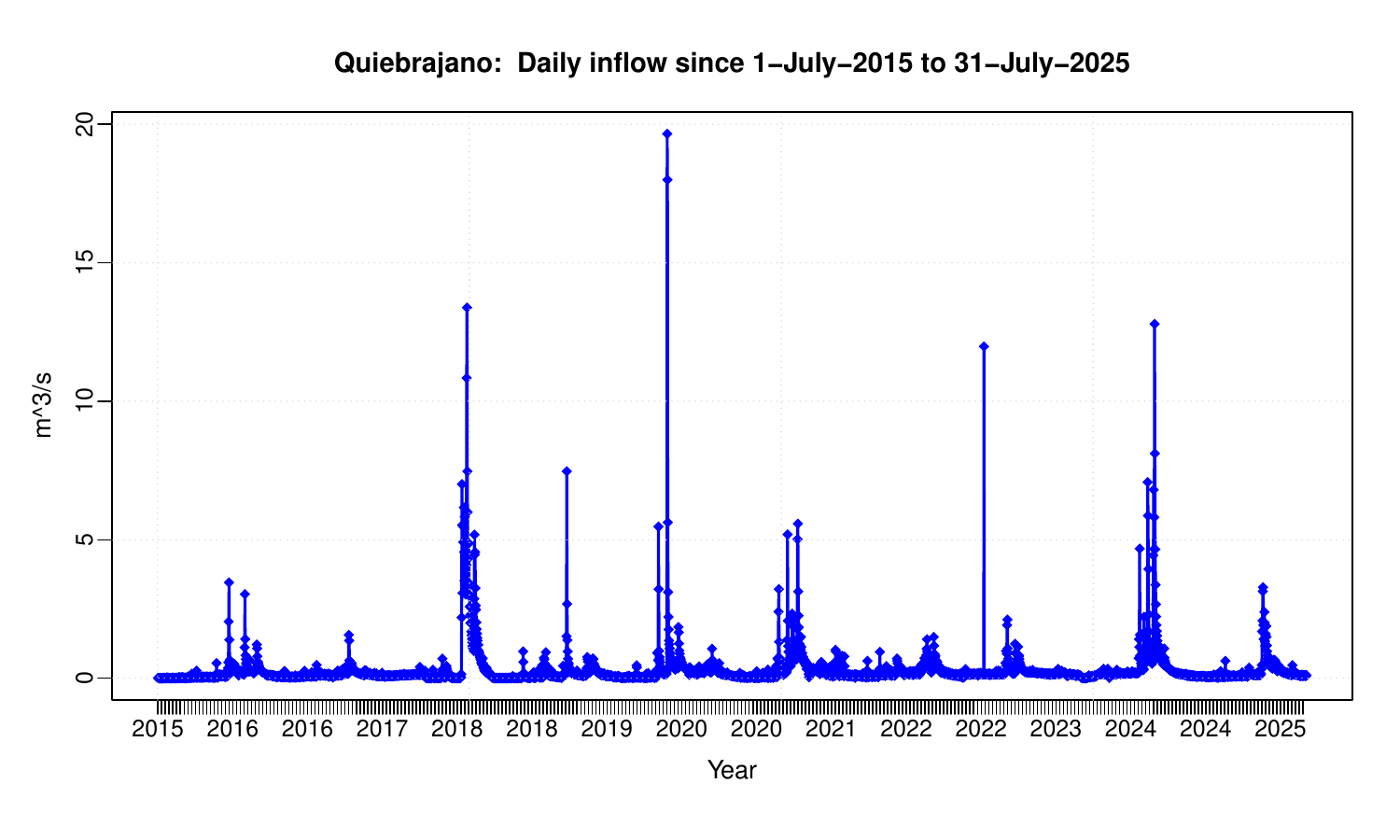}
\caption{Daily inflow (measured in $m^3/s$) recorded at the Quiebrajano Dam during the period 01/07/2015–31/07/2025 (data obtained from the SAIH System, accessed on 01/08/2025) }\label{fig:dam23}
\end{figure}

%
 In a very first approach, we have attempted to fit a PH-HMM to explain inflow patterns using daily observations the Quiebrajano pluviometric station  (Jaen, Spain). See Figure \ref{fig:dam23} for a graphical representation of daily inflow measured in $m^3/s$ recorded during the 10-year period from July-2015 to July-2025.

\begin{figure}[ht]
	\centering
	\includegraphics[width=0.8\textwidth]{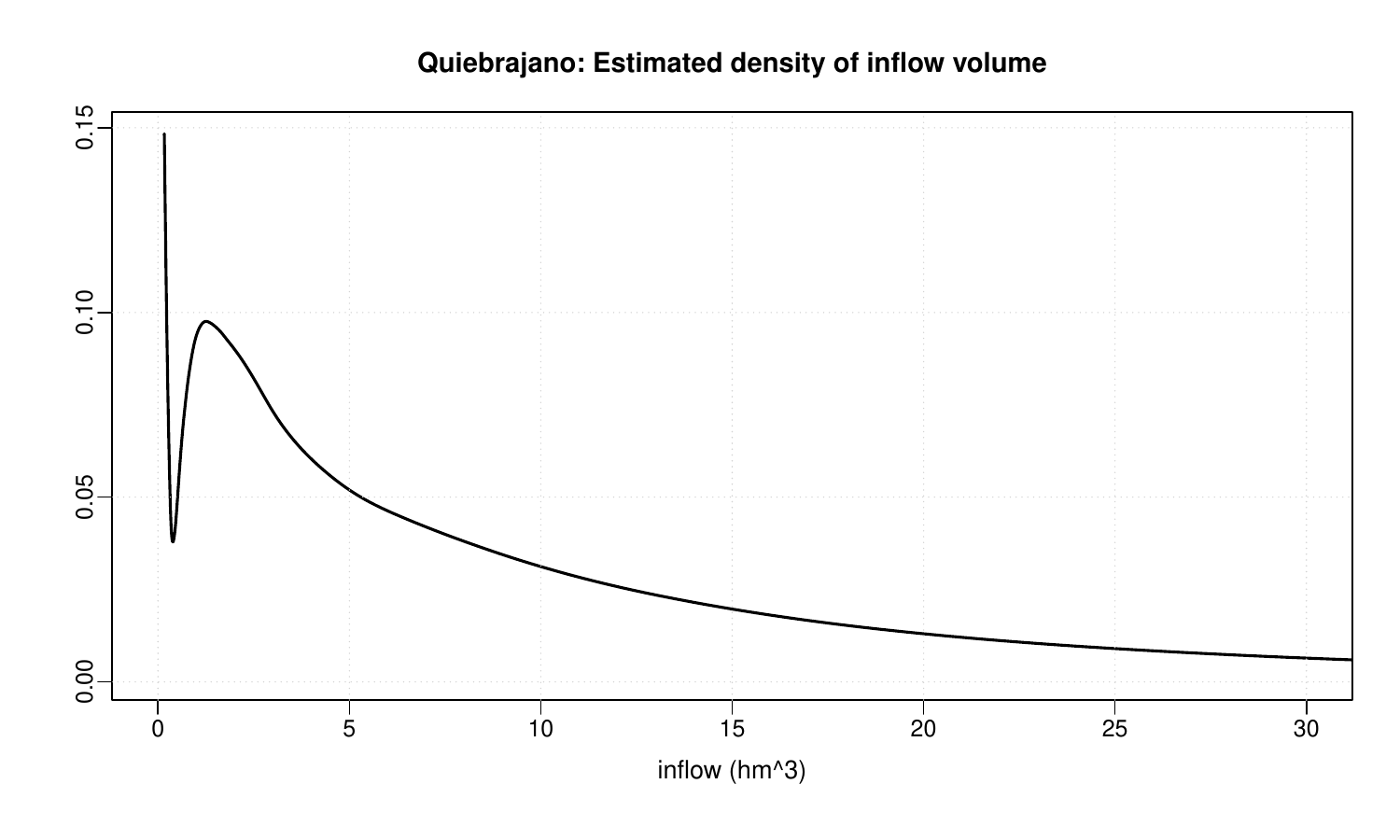}
	\caption{Density estimation of the annual inflow volume to the Quiebrajano Reservoir, recorded from October 1999 to October 2024.}
	\label{fig:densyear}
\end{figure}

\begin{figure}[ht]
	\centering
	\includegraphics[width=0.8\textwidth]{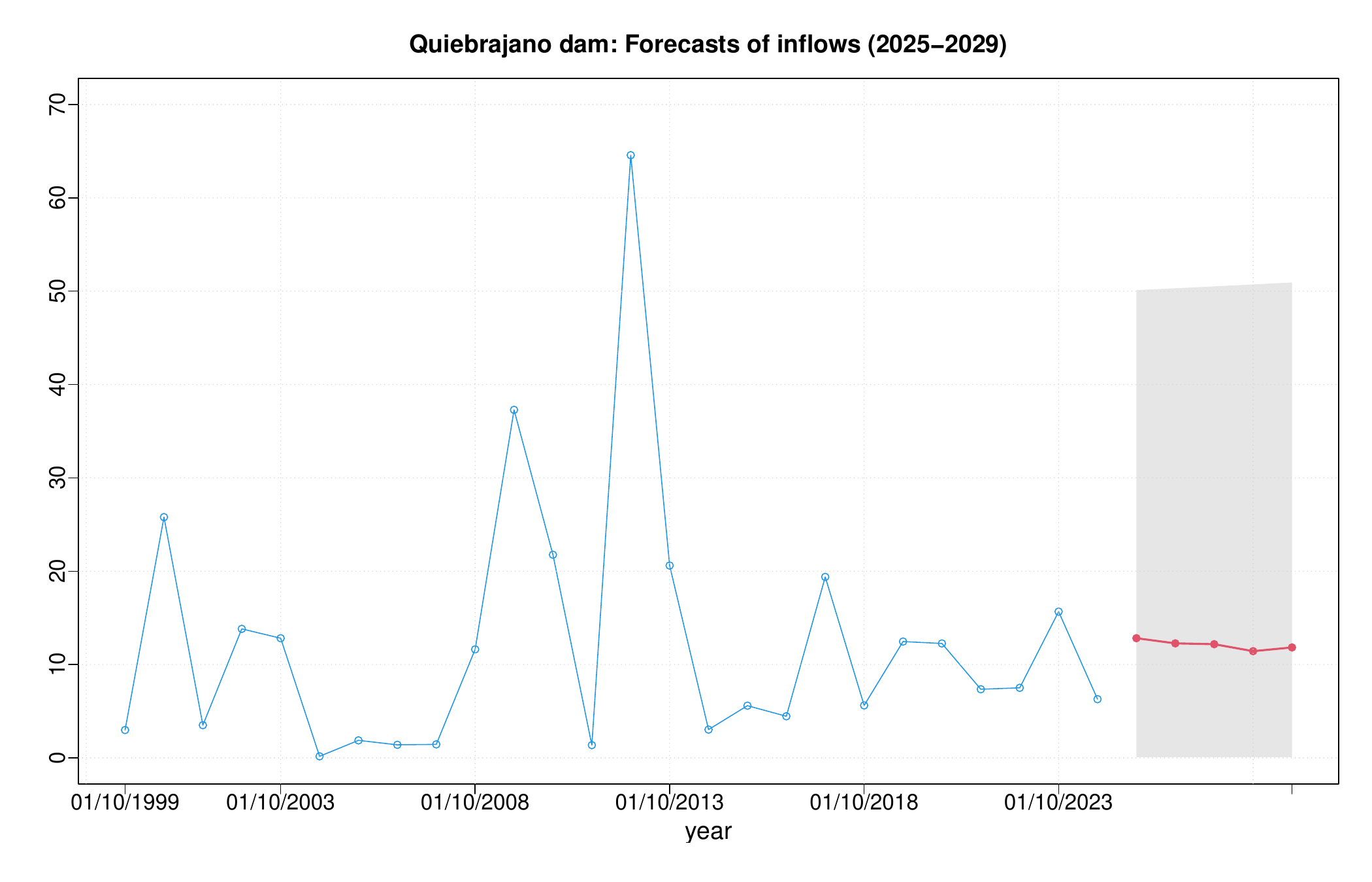}
	\caption{Predicted inflow values to the reservoir during the period 2025-2029 obtained using on the PH-HMM model }\label{fig:forecast}
	\end{figure}

A key finding after a preliminary inspection of the data, has been that a model with 3 hidden states might fit the data better than one with 2 states, suggesting 3 distinct rain/not-rain patterns observed in the data. In the numerical analysis presented in the following, we will consider yearly data. One key reason to use yearly data instead of monthly or daily data when predicting the state of a reservoir (such as stored volume or supply levels) is to reduce noise and focus on long-term trends. Daily and monthly data often contain high variability due to short-term weather events (like isolated storms or dry spells), measurement errors, or operational fluctuations. This can obscure underlying patterns and make predictive modelling more difficult. In contrast, annual data smooth out these short-term fluctuations, making it easier to identify long-term climatic trends, assess the impact of policy or management decisions, and build more robust models—especially when historical data is limited. For long-term planning, such as forecasting water availability over the next 5-year period (see Figure \ref{fig:forecast}, below), annual data often provide a clearer, more stable basis than high-frequency data.
We present in Figure \ref{fig:densyear} a nonparametric kernel density estimation based on observed annual inflow volumes, using the complete hydrological years available on the official website, starting on October 1$st$, 1999. The data covers a total of 26 years, up to September 30$th$, 2024. As shown in the graph, the inflow data exhibit clear heterogeneity, suggesting the presence of underlying regimes or states influenced by varying climatic or hydrological conditions. This justifies the use of a Hidden Markov Model (HMM), which is well-suited for modelling such latent state-dependent behaviour in time series \cite{Akintug2005}.

\bigskip

\noindent {\bf Results:}\\
We have considered 3 hidden states.
\begin{itemize}
	\item \textit{State 1}: The emission (inflow volume entering the reservoir) distribution under this state is modelled as an exponential distribution with rate $\lambda_1=6.006$. In other words, under this regime, the mean annual inflow to the reservoir is approximately 0.167 $hm^3$, which allows us to classify this regime as a {\bf drought} regime.

	\item \textit{State 2}: Under this regime, the inflow distribution is estimated as an exponential distribution with rate  $\lambda_{2}=0.626$, which corresponds to a mean annual inflow of approximately 1.653 $hm^3$. Compared to the unconditional mean inflow of around 22 $hm^3$ per year, this regime can be classified as {\bf dry}.

	\item \textit{State 3}: Under this regime, the inflow distribution is estimated as an exponential distribution with rate $\lambda_{3}=0.071$. The average annual inflow in this regime is approximately 14.08 $hm^3$, classifying it as a {\bf wet} regime. Notice that this value is very close to the annual inflow. 

\end{itemize}
The transition matrix corresponding to the extended state space HMM, is estimated as follows
\begin{equation}
\widehat{\widetilde{\bf P}}=
\begin{pmatrix}
 0.0000& 1.0000& 0.0000 &0.0000\\
 0.0000 &0.2651& 0.7349& 0.0000\\
 0.0000 &0.0000 &0.2254 &0.7746\\
0.0457 &0.0000& 0.0000 &0.9543
\end{pmatrix}
\end{equation}

The transition matrix of the embedded Markov chain in the hidden semi-Markov chain, with state space $E=\{\text{\bf drought, dry, wet}\}$,

\begin{equation}
	\widehat{\bf P}=
	\begin{pmatrix}
		0 & 1 &0 \\
		0 & 0 &1 \\
		1 & 0 &0
		\end{pmatrix}
\end{equation}

Finally it is obtained that the sojourn time in the hidden state {\bf dry} is distributed as a $\tt DPH$ with parameters
$\balpha_{2}=(1,0)$ and
\[
\widehat{{ T}}_{2}=	\begin{pmatrix}
0.2651& 0.7349\\
0.0000 &0.2254 
\end{pmatrix},
\]
which leads to estimate a mean sojourn time $\mu_{2}=\balpha_2 ({\bf I}-\widehat{T}_2)^{-1}{\bf e}=2.65$ years.

Finally, the sojourn time in state {\bf drought} is degenerate and equal to 1 year, and for the 
{\bf wet} regime we have a Geometric distribution with parameter 0.0457, which reports a mean sojourn time of 21.88 years.\\


\subsection{Building a Moran model to explain the volume of water stored in the dam}

\noindent{\bf Preliminary Review and Identification of Data Issues:}\\
\noindent Before delving into the case study of the Quiebrajano and developing the model using the available data, we would like to highlight certain inconsistencies identified on the SAIH website. These issues lead us to believe that expert assistance and complementary information are necessary to supplement the data provided online. Furthermore, in an ultimate case, it may be required to carry out corrective actions and maintenance on the website to ensure the accuracy and reliability of the information it offers.

Among the information available on the website, we can find the volume of water stored at the end of each hydrological year, which we define here as running from October 1 to September 30. We can also find the volume of inflow to the reservoir during each hydrological year. Using these values, along with the effective inflow to the reservoir each year, we can calculate the volume that should be stored in the reservoir at the beginning of each hydrological year simply by applying the equation that suits the Moran model and can be specified as follows, let $\widetilde{V}_0=V_0$ be the volume of water in the dam at 01-10-1999, which is the first record available on the website, and define
\[
\widetilde{V}_n=\min\{\max \{0,V_{n-1}+Y_{n-1}-O_{n-1}\}, C_1\}
\]
where
\begin{enumerate}
	\item $\widetilde{V}_n$ is the volume of water that should be stored at the beginning of the $n$th hydrological year, calculated according to the above balance equation;
	\item $Y_{n-1}$ is the total inflow during the year $n-1$;
	\item $O_{n-1}$ is the total outflow during the year $n-1$;
	\item $C_1$ is the maximum capacity of the dam.
\end{enumerate}

\begin{figure}[ht]
	\centering
	\includegraphics[width=0.9\textwidth]{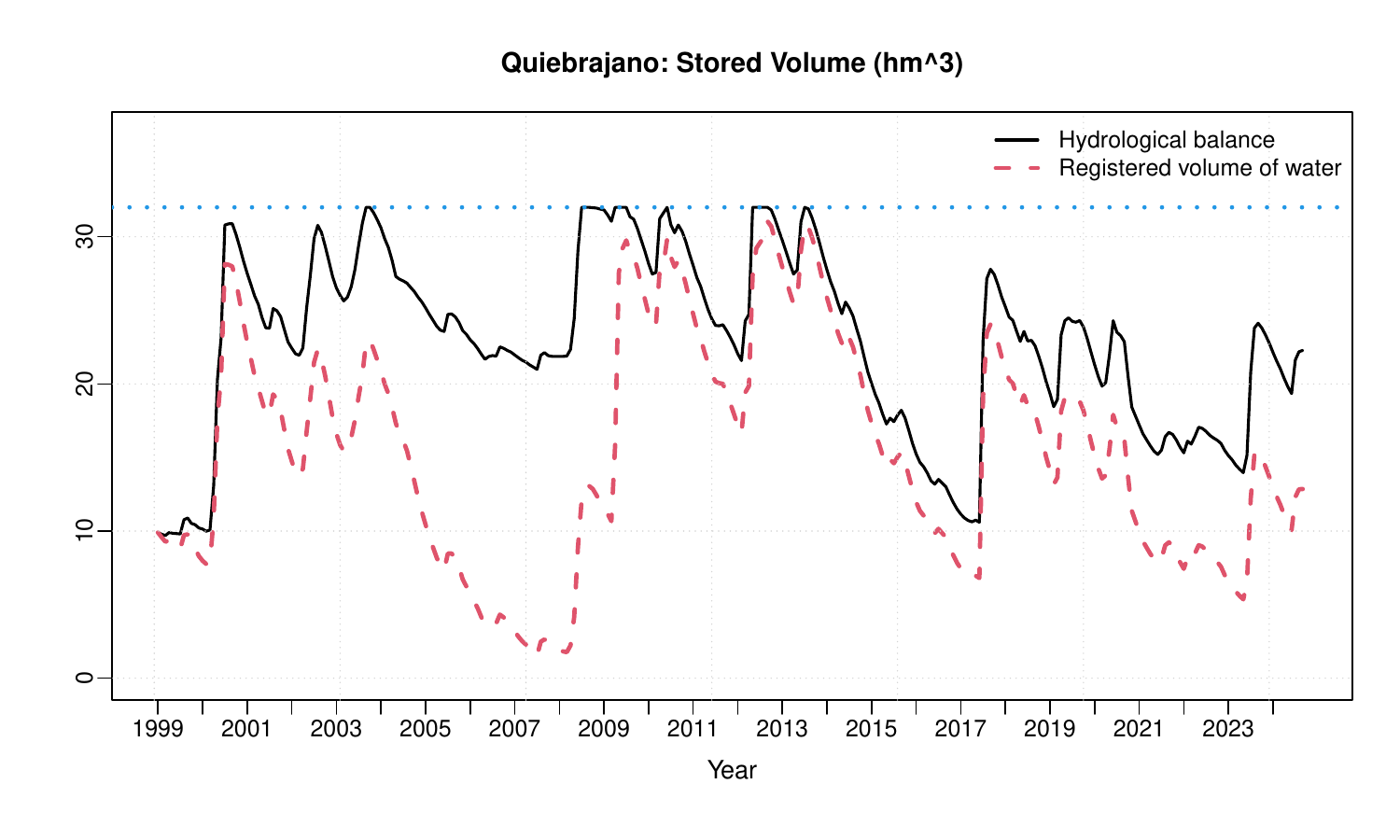}
	\caption{Volume of water stored per year from October-1999 to October-2024}\label{fig:yearvol}
	\end{figure}

The volume calculated year by year using this balance equation should be compared with the volume reported on the website in order to detect irregularities such as evaporation losses, infiltration from the reservoir to the ground, errors or inconsistencies in the data, water releases for agriculture, industry or other uses that are not officially recorded, underreponting or delays in updating values, etc. In our case we appreciate a big discrepancy at some years between the values of $\widetilde{V}_n$ and the registries obtained from the webpage, as can be seen in Figure \ref{fig:yearvol}.
	
\bigskip	
	
\noindent{\bf Building the model: }	\\
\noindent We now construct the Moran model based on the estimated transition matrix presented in the previous section (see equation~\eqref{eq:moran}), incorporating the specific characteristics of the Quiebrajano management system. The average annual released volume of water is estimated at $\omega = 10\, \text{hm}^3$. For illustrative purposes, we consider a simplified Moran model with four states, defined by the number of years during which the water demand can be met. Specifically, we consider: 0, 1, 2, and 3 or more years of supply.

Since we have fitted an Exponential distribution to the inflow variable $Y$, we adopt a discretization criterion based on the annual volume of water released. Specifically, we define zero inflow as occurring when $0 < Y < 1$. Based on this, the four states of the Moran model represent the number of consecutive years that water demand can be satisfied, even in the absence of additional inflow. The corresponding estimated transition matrix is given in equation \eqref{eq:morQui}.

	 \vspace{1cm}
	 \begin{equation}\label{eq:morQui}
	 	{\bf P}_{moran}=\begin{pmatrix}
	 	0.5769 & 0.2152 & 0.1057 & 0.1022 \\ 
	 	 0.1440 & 0.4329 & 0.2152 & 0.2079 \\ 
	 	0.0000 & 0.1440 & 0.4329 & 0.4231 \\ 
	 	0.0000 & 0.0000 & 0.1440 & 0.8560
	 	\end{pmatrix}
	 	\end{equation}
	
\bigskip
	 
\noindent{\bf Comparison with existing models}\\
\noindent Hydrological balance analyses are commonly conducted using time series methods, as these tools are well-suited for capturing temporal dependencies and trends in the data. To benchmark the performance of our proposed model, we compare it to a baseline based on an ARIMA model, a standard approach for modelling and forecasting hydrological time series (\cite{Salas1980}).

To evaluate the proposed PH-HMM, we compare it with a classical ARMA model fitted to the same annual inflow series. While ARMA models effectively capture short-term autocorrelations, they are limited in their ability to represent regime-dependent behaviours, such as prolonged droughts or wet periods. As a result, they deliver poorer estimation performance compared to our model.

Although ARIMA models are widely adopted in time series analysis, they perform poorly on the dataset considered in this study. For instance, automatic model selection using R’s auto.arima function suggests a trivial ARIMA(0,0,0) model, indicating that the series resembles white noise. More complex specifications, such as ARIMA(1,0,0), ARIMA(2,0,0), ARIMA(1,1,1), or ARIMA(0,0,2), not only yield higher AIC values but also fail to improve the model fit significantly (see Table \ref{tab:aic}). This illustrates the limitations of linear models when hidden state dynamics or nonlinear transitions are present in the data.

Given that the inflow series is strongly influenced by climate variability, it is reasonable to expect structural changes or regime shifts in the underlying process. Standard ARIMA models, being linear and limited in memory, are not capable of capturing such dynamics—explaining their underperformance in this context and reinforcing the need for alternative modelling approaches such as HMMs.

	\begin{table}[ht]
		\centering
		\begin{tabular}{|c|c|c|c|c|c|}
			\hline
			Model& AR(1)&AR(2)&MA(2)&Arima(1,1,1)&PH-MHH\\ \hline
			AIC &215.7773&217.6977&217.6396&211.9087&\textbf{204.3818}\\ \hline
			\end{tabular}\caption{Comparison of model PH-HMM with models ARIMA($p$,$d$,$q$) for different choices of parametres, in terms of the AIC statistics.}\label{tab:aic}
		\end{table}

\noindent{\bf Prediction of inflows}\\
\noindent Forecasted inflow values for the period 2025–2029 have been obtained based on the PH-HMM model estimated in the previous section. The plot includes predictions bands (shaded in grey) derived from 500 bootstrap samples generated using the fitted model. Additionally, the average values obtained across the 500 samples are shown as red dots. The blue curve represent the true data of inflows during the observations period. The bootstrap samples have been generated using the simulation algorithm explained in Section \ref{sec:measures}.

\section{Conclusions}

This paper introduced a Hidden Phase-Type Markov Model (HP-HMM) for modelling inflow processes to reservoirs under the influence of shifting weather regimes. The proposed framework extends conventional HMM approaches by incorporating general sojourn time distributions through phase-type modelling, allowing for more realistic representation of regime persistence and transitions. Furthermore, we presented a set of reliability measures derived from the model, relevant for reservoir management and risk assessment. Simulation results confirm the model's capacity to generate realistic inflow sequences and storage trajectories, supporting its potential use as a decision-support tool for water resource planning under uncertainty.

We applied the model to real data from a reservoir located in southern Spain, aiming to capture the dynamics of the hydrological balance under limited data availability. Discrepancies observed in official inflow records suggest the importance of expert validation in hydrological applications, especially when high-quality data are lacking.

In comparison with standard ARIMA models, our approach demonstrated superior performance in both statistical fit and interpretability, particularly in capturing latent regime shifts and nonlinear dynamics.

\end{document}